\documentclass[12pt,a4paper]{article}

\usepackage{ifthen} % for conditional statements
\newboolean{pdflatex}
\setboolean{pdflatex}{true} % False for eps figures 

\newboolean{articletitles}
\setboolean{articletitles}{true} % False removes titles in references

\newboolean{uprightparticles}
\setboolean{uprightparticles}{false} %True for upright particle symbols

\newboolean{inbibliography}
\setboolean{inbibliography}{false} %True once you enter the bibliography
\usepackage{multirow}

\def\paperauthors{LHCb collaboration} 
\def\paperasciititle{J/PSI and D0 production in SQRT(sNN)=68.5 GeV PbNe collisions} % Set ASCII title here
\def\papertitle{\jpsi and \Dz production in $\sqsnn =68.5\gev$ PbNe collisions }%} with LHCb in its fixed-target configuration}
\def\paperkeywords{{High Energy Physics}, {LHCb}} % Comma separated list
\def\papercopyright{2021 CERN for the benefit of the LHCb collaboration}
\def\paperlicence{CC-BY-4.0}
\def\paperlicenceurl{https://creativecommons.org/licenses/by/4.0/}

\usepackage[top=1in, bottom=1.25in, left=1in, right=1in]{geometry}
\usepackage{microtype}
\usepackage{lineno}  % for line numbering during review
\usepackage{xspace} % To avoid problems with missing or double spaces after
\usepackage{caption} %these three command get the figure and table captions automatically small

\usepackage{graphicx}  % to include figures (can also use other packages)
\usepackage{color}
\usepackage{colortbl}
\usepackage{amsmath} % Adds a large collection of math symbols
\usepackage{amssymb}
\usepackage{amsfonts}
\usepackage{upgreek} % Adds in support for greek letters in roman typeset

\newcommand*\patchAmsMathEnvironmentForLineno[1]{%
\expandafter\let\csname old#1\expandafter\endcsname\csname #1\endcsname
\expandafter\let\csname oldend#1\expandafter\endcsname\csname
end#1\endcsname
 \renewenvironment{#1}%
   {\linenomath\csname old#1\endcsname}%
   {\csname oldend#1\endcsname\endlinenomath}%
}
\newcommand*\patchBothAmsMathEnvironmentsForLineno[1]{%
  \patchAmsMathEnvironmentForLineno{#1}%
  \patchAmsMathEnvironmentForLineno{#1*}%
}
\AtBeginDocument{%
\patchBothAmsMathEnvironmentsForLineno{equation}%
\patchBothAmsMathEnvironmentsForLineno{align}%
\patchBothAmsMathEnvironmentsForLineno{flalign}%
\patchBothAmsMathEnvironmentsForLineno{alignat}%
\patchBothAmsMathEnvironmentsForLineno{gather}%
\patchBothAmsMathEnvironmentsForLineno{multline}%
\patchBothAmsMathEnvironmentsForLineno{eqnarray}%
}

\usepackage{hyperxmp}

\usepackage[pdftex,
            pdfauthor={\paperauthors},
            pdftitle={\paperasciititle},
            pdfkeywords={\paperkeywords},
            pdfcopyright={Copyright (C) \papercopyright},
            pdflicenseurl={\paperlicenceurl}]{hyperref}

\usepackage[all]{hypcap} % Internal hyperlinks to floats.

\usepackage{xspace} 
\usepackage{upgreek}

\def\lhcb   {\mbox{LHCb}\xspace}

\def\MagUp {\mbox{\em Mag\kern -0.05em Up}\xspace}

\ifthenelse{\boolean{uprightparticles}}%
{

 \def\Pmu         {\ensuremath{\upmu}\xspace}

 \def\Ppi         {\ensuremath{\uppi}\xspace}

 \def\Ppsi        {\ensuremath{\uppsi}\xspace}

 \def\PDelta      {\ensuremath{\Delta}\xspace}                 
 \def\PXi         {\ensuremath{\Xi}\xspace}                 
 \def\PLambda     {\ensuremath{\Lambda}\xspace}                 
 \def\PSigma      {\ensuremath{\Sigma}\xspace}                 
 \def\POmega      {\ensuremath{\Omega}\xspace}                 
 \def\PUpsilon    {\ensuremath{\Upsilon}\xspace}
 \let\oldPi\Pi
 \def\PPi         {\ensuremath{\oldPi}\xspace}

 \def\PB      {\ensuremath{\mathrm{B}}\xspace}                 
                  
 \def\PD      {\ensuremath{\mathrm{D}}\xspace}

 \def\PJ      {\ensuremath{\mathrm{J}}\xspace}                 
 \def\PK      {\ensuremath{\mathrm{K}}\xspace}

 \def\Pb      {\ensuremath{\mathrm{b}}\xspace}                 
 \def\Pc      {\ensuremath{\mathrm{c}}\xspace}

 \def\Pi      {\ensuremath{\mathrm{i}}\xspace}

 \def\Ps      {\ensuremath{\mathrm{s}}\xspace}

 \def\thebaroffset{0.0em}
}
{

 \def\Pmu         {\ensuremath{\mu}\xspace}

 \def\Ppi         {\ensuremath{\pi}\xspace}

 \def\Ppsi        {\ensuremath{\psi}\xspace}                 
                  
 \mathchardef\PDelta="7101
 \mathchardef\PXi="7104
 \mathchardef\PLambda="7103
 \mathchardef\PSigma="7106
 \mathchardef\POmega="710A
 \mathchardef\PUpsilon="7107
 \mathchardef\PPi="7105
                  
 \def\PB      {\ensuremath{B}\xspace}                 
                  
 \def\PD      {\ensuremath{D}\xspace}

 \def\PJ      {\ensuremath{J}\xspace}                 
 \def\PK      {\ensuremath{K}\xspace}

 \def\Pb      {\ensuremath{b}\xspace}                 
 \def\Pc      {\ensuremath{c}\xspace}

 \def\Pi      {\ensuremath{i}\xspace}

 \def\Ps      {\ensuremath{s}\xspace}

 \def\thebaroffset{0.18em}
}
\newcommand{\offsetoverline}[2][\thebaroffset]{\kern #1\overline{\kern -#1 #2}}%

\makeatletter
\ifcase \@ptsize \relax% 10pt
  \newcommand{\miniscule}{\@setfontsize\miniscule{4}{5}}% \tiny: 5/6
\or% 11pt
  \newcommand{\miniscule}{\@setfontsize\miniscule{5}{6}}% \tiny: 6/7
\or% 12pt
  \newcommand{\miniscule}{\@setfontsize\miniscule{5}{6}}% \tiny: 6/7
\fi
\makeatother

\DeclareRobustCommand{\optbar}[1]{\shortstack{{\miniscule (\rule[.5ex]{1.25em}{.18mm})}
  \\ [-.7ex] $#1$}}

\def\mup        {{\ensuremath{\Pmu^+}}\xspace}
\def\mun        {{\ensuremath{\Pmu^-}}\xspace} % muon negative (\mum is taken)

\def\squark    {{\ensuremath{\Ps}}\xspace}

\def\cquark    {{\ensuremath{\Pc}}\xspace}
\def\cquarkbar {{\ensuremath{\overline \cquark}}\xspace}
\def\ccbar     {{\ensuremath{\cquark\cquarkbar}}\xspace}
\def\bquark    {{\ensuremath{\Pb}}\xspace}

\def\pion   {{\ensuremath{\Ppi}}\xspace}

\def\pipm   {{\ensuremath{\pion^\pm}}\xspace}

\def\kaon    {{\ensuremath{\PK}}\xspace}
\def\KorKbar {\kern \thebaroffset\optbar{\kern -\thebaroffset \PK}{}\xspace}

\def\Kmp     {{\ensuremath{\kaon^\mp}}\xspace}

\def\D       {{\ensuremath{\PD}}\xspace}

\def\DorDbar {\kern \thebaroffset\optbar{\kern -\thebaroffset \PD}\xspace}
\def\Dz      {{\ensuremath{\D^0}}\xspace}

\def\Dp      {{\ensuremath{\D^+}}\xspace}
\def\Dm      {{\ensuremath{\D^-}}\xspace}

\def\DpDm    {\ensuremath{\Dp {\kern -0.16em \Dm}}\xspace}

\def\B       {{\ensuremath{\PB}}\xspace}

\def\BorBbar {\kern \thebaroffset\optbar{\kern -\thebaroffset \PB}\xspace}

\def\Bd      {{\ensuremath{\B^0}}\xspace}

\def\BdorBdbar {\kern \thebaroffset\optbar{\kern -\thebaroffset \Bd}\xspace}

\def\Bs      {{\ensuremath{\B^0_\squark}}\xspace}

\def\BsorBsbar {\kern \thebaroffset\optbar{\kern -\thebaroffset \Bs}\xspace}

\def\jpsi     {{\ensuremath{{\PJ\mskip -3mu/\mskip -2mu\Ppsi}}}\xspace}

\def\Y#1S{\ensuremath{\PUpsilon{(#1S)}}\xspace}

\def\LorLbar     {\kern \thebaroffset\optbar{\kern -\thebaroffset \PLambda}\xspace}

\def\BF         {{\ensuremath{\mathcal{B}}}\xspace}

\def\to                 {\ensuremath{\rightarrow}\xspace}

\def\AT#1     {\ensuremath{A_{\mathrm{T}}^{#1}}\xspace}           % 2

\def\C#1      {\ensuremath{\mathcal{C}_{#1}}\xspace}                       % 9
\def\Cp#1     {\ensuremath{\mathcal{C}_{#1}^{'}}\xspace}                    % 7
\def\Ceff#1   {\ensuremath{\mathcal{C}_{#1}^{\mathrm{(eff)}}}\xspace}        % 9  
\def\Cpeff#1  {\ensuremath{\mathcal{C}_{#1}^{'\mathrm{(eff)}}}\xspace}       % 7
\def\Ope#1    {\ensuremath{\mathcal{O}_{#1}}\xspace}                       % 2
\def\Opep#1   {\ensuremath{\mathcal{O}_{#1}^{'}}\xspace}                    % 7

\newcommand{\aunit}[1]{\ensuremath{\text{\,#1}}}       
\newcommand{\tev}{\aunit{Te\kern -0.1em V}\xspace}
\newcommand{\gev}{\aunit{Ge\kern -0.1em V}\xspace}
\newcommand{\mev}{\aunit{Me\kern -0.1em V}\xspace}
\newcommand{\kev}{\aunit{ke\kern -0.1em V}\xspace}
\newcommand{\ev}{\aunit{e\kern -0.1em V}\xspace}
 
\newcommand{\mevc}{\ensuremath{\aunit{Me\kern -0.1em V\!/}c}\xspace}
\newcommand{\gevc}{\ensuremath{\aunit{Ge\kern -0.1em V\!/}c}\xspace}
\newcommand{\mevcc}{\ensuremath{\aunit{Me\kern -0.1em V\!/}c^2}\xspace}
\newcommand{\gevcc}{\ensuremath{\aunit{Ge\kern -0.1em V\!/}c^2}\xspace}
\newcommand{\stat}{\aunit{(stat.)}\xspace}
\newcommand{\syst}{\aunit{(syst.)}\xspace}

\def\gsim{{~\raise.15em\hbox{$>$}\kern-.85em
          \lower.35em\hbox{$\sim$}~}\xspace}
\def\lsim{{~\raise.15em\hbox{$<$}\kern-.85em
          \lower.35em\hbox{$\sim$}~}\xspace}

\def\sqsnn {\ensuremath{\protect\sqrt{s_{\scriptscriptstyle\text{NN}}}}\xspace}
\def\pt         {\ensuremath{p_{\mathrm{T}}}\xspace}

\def\evtgen     {\mbox{\textsc{EvtGen}}\xspace}

\def\geant      {\mbox{\textsc{Geant4}}\xspace}

\def\photos     {\mbox{\textsc{Photos}}\xspace}

\def\pythia     {\mbox{\textsc{Pythia}}\xspace}

\def\tell1  {TELL1\xspace}
\def\ukl1   {UKL1\xspace}

\newcommand{\lhcborcid}[1]{\href{https://orcid.org/#1}{\hspace*{0.1em}\raisebox{-0.45ex}{\includegraphics[width=1em]{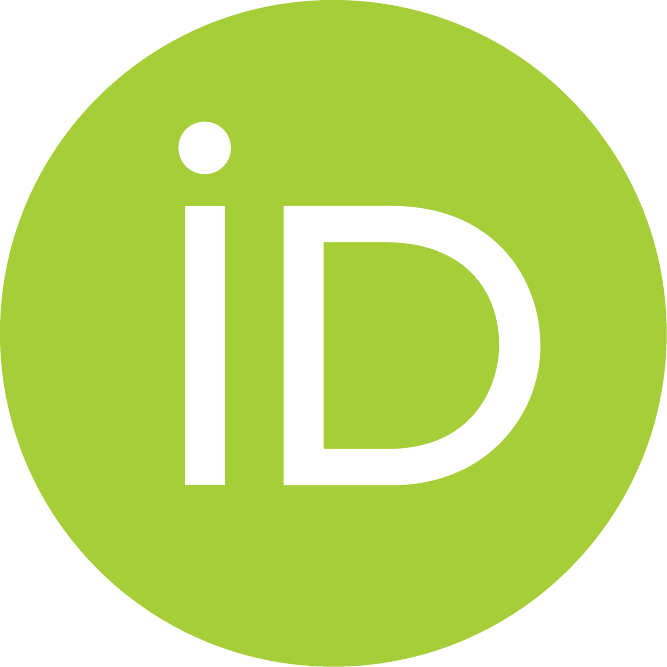}}}}

\def\epos      {\mbox{\textsc{Epos}}\xspace}
\newcommand{\pp}{\ensuremath{pp}\xspace}

\newcommand{\pNe}{\ensuremath{p\text{Ne}}\xspace}
\def\ncoll	{\ensuremath{N_{\mathrm{coll}}}\xspace}
\def\mncoll	{\ensuremath{\langle N_{\mathrm{coll}}\rangle}\xspace}

\usepackage{cite} % Allows for ranges in citations
\usepackage{mciteplus}
\usepackage{longtable} % only for template; not usually to be used in PAPERs

\begin{document}

\renewcommand{\thefootnote}{\fnsymbol{footnote}}
\setcounter{footnote}{1}

%%%%%%%%%%%%%%%%%%%%%%%%%
%%%%%  TITLE PAGE  %%%%%%
%%%%%%%%%%%%%%%%%%%%%%%%%
\begin{titlepage}
\pagenumbering{roman}
% Header ---------------------------------------------------
\vspace*{-1.5cm}
\centerline{\large EUROPEAN ORGANIZATION FOR NUCLEAR RESEARCH (CERN)}
\vspace*{1.5cm}
\noindent
\begin{tabular*}{\linewidth}{lc@{\extracolsep{\fill}}r@{\extracolsep{0pt}}}
\ifthenelse{\boolean{pdflatex}}% Logo format choice
{\vspace*{-1.5cm}\mbox{\!\!\!\includegraphics[width=.14\textwidth]{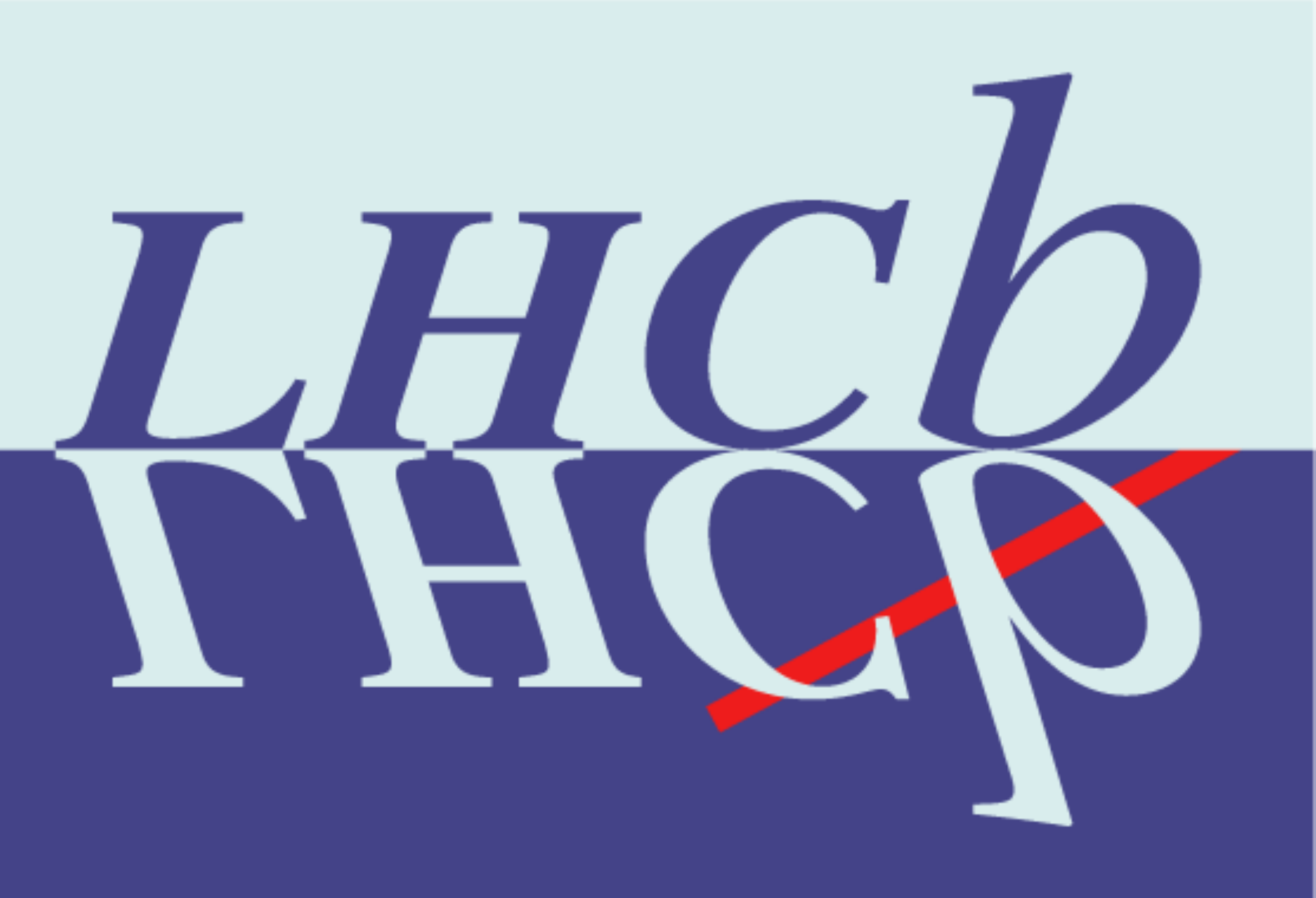}} & &}%
{\vspace*{-1.2cm}\mbox{\!\!\!\includegraphics[width=.12\textwidth]{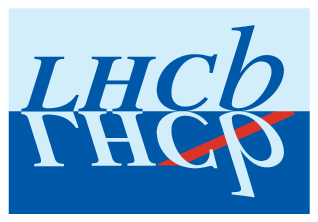}} & &}%
\\
 & & CERN-EP-2022-215 \\  % ID 
 & & LHCb-PAPER-2022-011  \\  % ID 
 & & August 11, 2023 \\ % Date - Can also hardwire e.g.: 23 March 2010
 & & \\
% not in paper \hline
\end{tabular*}
\vspace*{4.0cm}
% Title --------------------------------------------------
{\normalfont\bfseries\boldmath\huge
\begin{center}
% DO NOT EDIT HERE. Instead edit macro in main.tex to keep metadata correct
  \papertitle 
\end{center}
}
\vspace*{2.0cm}
% Authors -------------------------------------------------
\begin{center}
%In the footnote, replace 'paper' by 'Letter' in case of submission to PRL or PLB 
% Edit macro in main.tex to keep metadata correct
\paperauthors\footnote{Authors are listed at the end of this paper.}
\end{center}
\vspace{\fill}
% Abstract -----------------------------------------------
\begin{abstract}
  \noindent
  
The first measurement of \jpsi and \Dz production in PbNe collisions by the LHCb experiment in its fixed-target configuration is reported. The production of \jpsi and \Dz mesons is studied with a beam of lead ions with an energy of 2.5\tev per nucleon  colliding on gaseous neon targets at rest, corresponding to a nucleon-nucleon centre-of-mass energy of $\sqsnn=68.5\gev$. The \jpsi/\Dz production cross-section ratio is studied as a function of rapidity, transverse momentum and collision centrality. 
These data are compared with measurements from \pNe collisions at the same energy and show no difference in the observed \jpsi suppression trend when comparing \pNe and PbNe peripheral collisions with PbNe central collisions.
\end{abstract}
\vspace*{2.0cm}
\begin{center}
Published in Eur.~Phys.~J.~\textbf{C83} (2023) 658. 
\end{center}
\vspace{\fill}
{\footnotesize 
% Edit macro in main.tex to keep metadata correct
\centerline{\copyright~\papercopyright. \href{\paperlicenceurl}{\paperlicence} licence.}}
\vspace*{2mm}
\end{titlepage}

%%%%%%%%%%%%%%%%%%%%%%%%%%%%%%%%
%%%%%  EOD OF TITLE PAGE  %%%%%%
%%%%%%%%%%%%%%%%%%%%%%%%%%%%%%%%

%  empty page follows the title page ----
\newpage
\setcounter{page}{2}
\mbox{~}
\cleardoublepage
\renewcommand{\thefootnote}{\arabic{footnote}}
\setcounter{footnote}{0}

%%%%%%%%%%%%%%%%%%%%%%%%%
%%%%% Main text %%%%%%%%%
%%%%%%%%%%%%%%%%%%%%%%%%%

\pagestyle{plain} % restore page numbers for the main text
\setcounter{page}{1}
\pagenumbering{arabic}

%\linenumbers

%%%%% Introduction 
In the high-density and high-temperature regime of quantum chromodynamics (QCD), the production of heavy quarks in nucleus-nucleus interactions is well suited to study the transition between ordinary hadronic matter and the hot and dense quark-gluon plasma (QGP). Since heavy quark masses are large compared to the QGP critical temperature, $T_c\sim 156\mev$ \cite{QGPTc}, their production occurs in primary nucleon-nucleon collisions, at an early stage of the interaction. They can therefore experience the full evolution of the created nuclear medium. Moreover, at sufficiently high temperature, larger than $T_c$, lattice QCD calculations predict that the production of  charmonium (\ccbar bound states) decreases with respect to the overall \ccbar production, due to the modification of their binding mechanism \cite{Digal}. 
Consequently, the proper interpretation of the charmonium suppression, observed in nucleus-nucleus collisions at various energies \cite{KluSa,Review}, requires the measurement of charmonium yields together with the overall charm quark production \cite{Satz}. 
Since most of the charm quarks hadronise into open charm \Dz mesons, the \Dz production yield should provide a suitable reference for the study of the charmonium yield modification, assuming that \Dz production is not modified by the medium.

In this paper, the first measurement of \jpsi and \Dz production in fixed-target nucleus-nucleus collisions at the LHC is presented.\footnote{The inclusion of charge-conjugate processes is implied throughout the paper.}
Lead ions with an energy per nucleon of 2.5~TeV are incident on neon nuclei at rest, corresponding to a nucleon-nucleon centre-of-mass energy of $\sqrt{s_{NN}}=68.5\gev$. The \jpsi and \Dz mesons are measured via their \mup\mun and $K^\mp \pi^\pm$ decays, respectively. 

%%%% Detector
The LHCb detector~\cite{Alves:2008zz, LHCb-DP-2014-002} is a single-arm forward spectrometer covering the pseudorapidity range $2 < \eta < 5$. It has been designed for the study of particles containing \cquark or \bquark quarks. The main detector elements are: the silicon-strip vertex locator (VELO) surrounding the interaction region that allows \cquark and \bquark hadrons to be identified from their characteristic flight distance; a tracking system that provides a measurement of the momentum of charged particles, with a relative uncertainty that varies from $0.5\%$ at low momentum to $1.0\%$ at 200 \gev; two ring-imaging Cherenkov detectors that are able to discriminate among different types of charged hadrons; a calorimeter system consisting of scintillating-pad and preshower detectors, electromagnetic and hadronic calorimeters; and a muon detector composed of alternating layers of iron and multiwire proportional chambers. 
The system for measuring the overlap with gas (SMOG)~\cite{LHCb-PAPER-2014-047}, which enables the injection of gases with pressures of $O(10^{-7})$ mbar in the beam pipe section inside the VELO, is used to operate LHCb as a fixed-target experiment. The SMOG allows the injection of noble gases and therefore gives the unique opportunity to
study nucleus-nucleus and proton-nucleus collisions with various targets. Due to the boost of the centre-of-mass frame induced by the high-energy lead ion beam, the LHCb acceptance extends over the backward
rapidity hemisphere in the centre-of-mass system of the reaction covering the region $-2.29 \lesssim y^\star \lesssim 0$. 

%%%% Trigger
Selected events are recorded by a two-stage trigger system \cite{trigger}. The  first level is implemented in hardware and uses information provided by the calorimeters and the muon detectors, while the second is a software trigger. The hardware trigger requires at least one identified muon for the selection of the $\jpsi \to \mup \mun$ decay and energy deposit in the calorimeter for the $\Dz \to K^\mp \pi^\pm$ selection. The software trigger requires two well-reconstructed muons for the \jpsi mode, while for the \Dz channel, the software trigger requires a track with transverse momentum larger than 500\mevc.  
 
%%%%% Data
The data samples are collected under dedicated beam conditions where lead ion bunches moving towards the detector do not cross any bunch moving in the opposite direction at the nominal lead-lead (PbPb) interaction point. Events must have a reconstructed primary vertex (PV) $z$ position (along the beam direction) within the fiducial region $z_{PV}\in[-200, 150]$ mm where high reconstruction efficiencies are achieved and calibration samples are available. In order to suppress residual PbPb collisions occuring at $z_{PV} = 0$ mm, events with activity in the backward region, with respect to the PV position, are vetoed, based on the number of hits in the VELO stations upstream of the interaction region. 

%%%%% Offline selection
The offline selection of \jpsi and \Dz candidates is similar to that used in Ref.~\cite{LHCb-PAPER-2018-023}. Events must contain a primary vertex with at least four tracks reconstructed in the VELO detector.  The \jpsi candidates are formed from two oppositely charged muons forming a vertex. The muons must have transverse momentum, \pt, larger than 500 MeV$/c$ and are required to be consistent with originating from the primary interaction point. Each \Dz candidate is formed from an oppositely charged pair of charged particles that come from a common displaced vertex. One of these particles must be identified as a kaon and the other as a pion. The \Dz candidates are required to have a decay time larger than 0.4 ps. The measurements are performed in the range of \Dz and \jpsi transverse momentum $\pt < 8\gevc$ and rapidity $2.0 < y < 4.29$.

%%%%%% Efficiency
The kinematic acceptance and reconstruction efficiencies are determined using simulated lead-neon (PbNe) collision events. 
In the simulation, \jpsi and \Dz mesons are generated using \pythia~8 \cite{Pythia8} with a specific LHCb configuration \cite{1742-6596-331-3-032047}  and with colliding lead-ion beam momenta equal to the momentum per nucleon of the beam and target in the centre-of-mass frame. Particle decays are described by \evtgen \cite{EvtGen}, in which final-state radiation is generated using \photos \cite{PHOTOS}. The four-momenta of the \jpsi and \Dz decay products are embedded into PbNe minimum bias events that are generated with the \epos event generator \cite{Pierog:2013ria}  using beam parameters obtained from  data. Decays of hadronic particles generated with \epos are also described by \evtgen. The interaction of the generated particles with the detector, and its response, are implemented using the \geant toolkit \cite{geant2, geant4} as described in Ref. \cite{gauss}. After reconstruction, the simulated events are assigned weights using a gradient boosted reweighter \cite{Rogozhnikov:2016bdp} to reproduce the data with respect to the multiplicity in the scintillating pad detector (denoted hereafter as hit multiplicity) and the transverse momentum distribution of the \jpsi and \Dz candidates. The overall efficiencies, including the kinematic acceptance, the efficiencies of the trigger, event selection, primary vertex reconstruction, tracking and particle identification, are 13.9$\%$ and 1.1$\%$ for the \jpsi and  \Dz decays, respectively.

\begin{figure}[t]
\begin{center}
\includegraphics[width=0.49\textwidth]{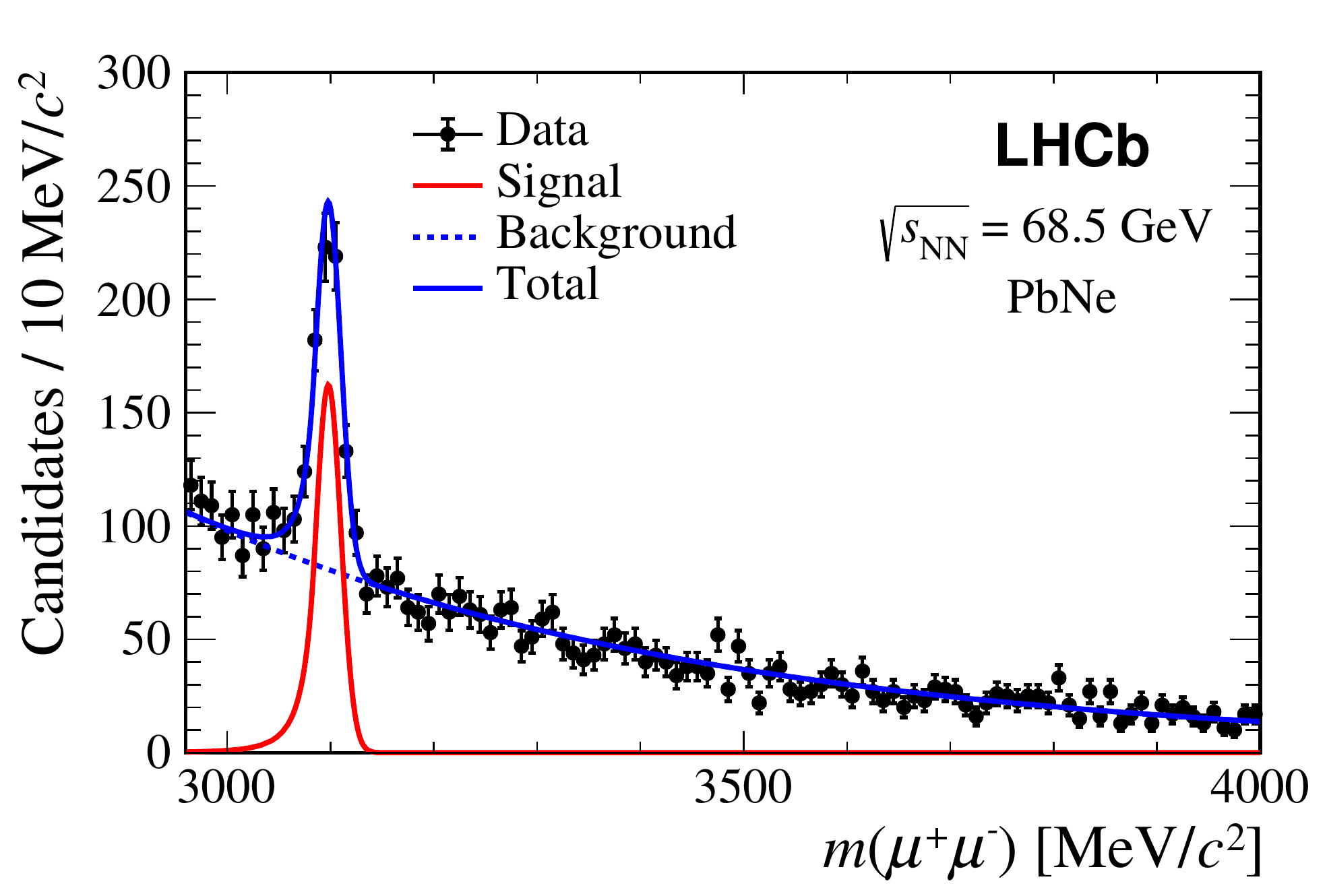}
\includegraphics[width=0.49\textwidth]{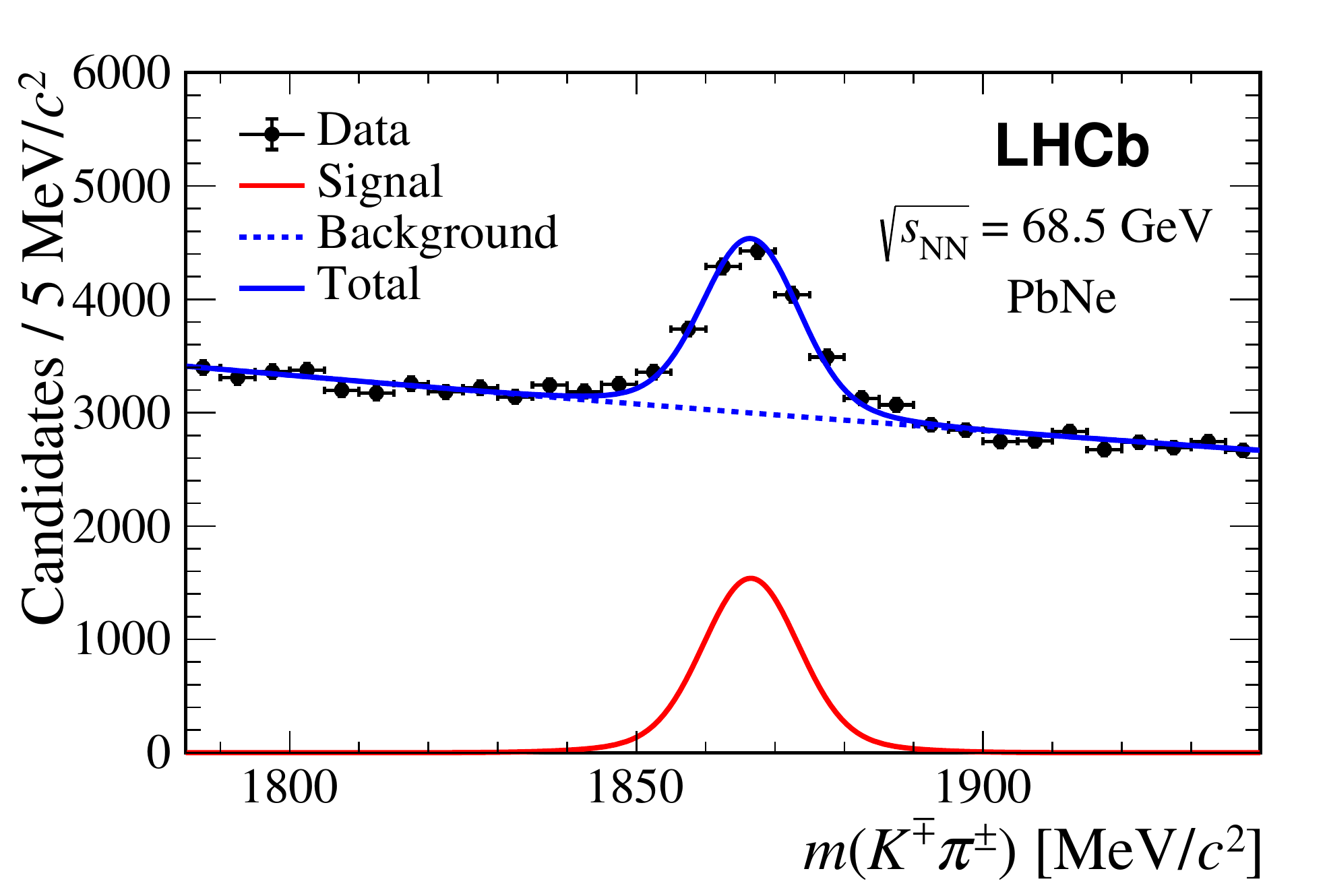}
\caption{Mass distributions of  (left) $\jpsi\to\mun\mup$ and (right) $\Dz\to\Kmp\pipm$  candidates. The data are overlaid with the fit function.}
\label{fig::InvMass}
\end{center}
\end{figure}

The \jpsi and \Dz signal yields are obtained from extended unbinned maximum-likelihood fits to their mass distributions. The \jpsi signal is described by a Crystal Ball function\cite{CB} while the \Dz signal is described by the sum of two Gaussian functions. The background components are modelled by exponential functions. Figure \ref{fig::InvMass} shows the mass distribution obtained after all selection criteria are applied to the entire PbNe data set, with the fit functions superimposed. The overall yields for the \jpsi and \Dz channels are about $550$ and $5700$ respectively. 
The signal yields are determined in intervals of $\pt$, $y^{\star}$ and hit multiplicity. The yields determined from the mass fit are corrected for the geometrical acceptance of the detector and the efficiencies of the trigger, event selection, primary vertex reconstruction, tracking and particle identification. The efficiencies for particle identification \cite{LHCb-PUB-2016-021} and tracking are obtained from a control sample of lead-proton (Pb$p$) collision data. All the other efficiencies are determined from simulation. Several sources of systematic uncertainties are considered, affecting either the determination of the signal yields or the total efficiencies. They are summarised in Table~\ref{tab:systematics} separately for correlated and uncorrelated systematic uncertainties among the bins.

\begin{table}[t]
\caption{\small \label{tab:systematics} Summary of the systematic uncertainties affecting the \jpsi/\Dz production ratio. Those that are correlated between bins affect all measurements by the same relative amount. Ranges denote the minimum and the maximum values among the $y$, \pt or hit multiplicity bins. }
\begin{center}
{\small
\begin{tabular}{@{}lc@{}}
\hline
{\textbf{Systematic uncertainties}} &  \jpsi/\Dz\\
\hline
{\textbf{Correlated bin uncertainties}} & \\
Simulation weighting & $16.0\%$\\
MC-truth matching efficiency &  $4.1\%$ \\
PV efficiency &   $0.3\%$ \\
Tracking efficiency                &  $2.9\%$ \\
Particle identification & $1.2\%$ \\
Overall \jpsi/\Dz determination & $7.8\%$ \\
% Average value  & $5.3\%$ \\
{\textbf{Uncorrelated bin uncertainties}} & \\
Signal mass model                 & $[1.1, 14.7]\%$ \\
Simulation statistical error                    &  $[0.5, 1.8]\%$ \\
Tracking calibration sample stat.    & $<0.1\%$ \\
Particle identification calibration stat.   & $[6.5, 6.9]\%$ \\
\hline
\textbf{Overall systematic uncertainty} & $19.6\%$\\
\hline

\end{tabular} 
}
\end{center}
\end{table}

The contamination from residual PbPb collisions, evaluated to be 0.23$\%$, is considered negligible in the \jpsi/\Dz production ratio. The fraction of signal from \bquark-hadron decays at $\sqsnn = 68.5$ GeV is estimated to be less than 0.1$\%$, which is also considered to be negligible.
The PV efficiency systematic uncertainty is determined by considering the difference between the PV efficiency computed using the \jpsi and \Dz simulations.
The systematic uncertainty related to the mass fit is evaluated using alternative models for signal and background shapes that reproduce the mass shapes  well. Another source of uncertainty arises from the finite size of the simulation sample used to compute the acceptances and efficiencies. Systematic uncertainties in the tracking and particle identification efficiencies are mainly related to the differences between the track multiplicity in PbNe and Pb$p$ collisions, and to the size of the Pb$p$ samples. The tracking systematic uncertainty also takes into account the difference in tracking efficiency between the data and the simulation. The MC-truth matching procedure is also tested and its inefficiency is considered a systematic uncertainty. Alternative sets of event weights  are produced to evaluate the uncertainty associated with the reweighting procedure. The main variation is obtained  by different random splitting of the training/testing samples and with different multiplicity input variables (such as the number of long tracks, or the multiplicity hits, or the clusters or tracks in the VELO). The largest discrepancy between the data and simulation is in the  primary vertex position distribution. This discrepancy mostly cancels in the \jpsi/\Dz ratio. To obtain the residual discrepancy, the difference between the overall \jpsi/\Dz efficiency ratio evaluated using the default weights and the same ratio computed with the alternative weights is studied with respect to the primary vertex position, leading to another systematic uncertainty affecting the weighting. These tests result in a correlated 16$\%$ systematic uncertainty to be applied in the $J/\psi/D^0$ production ratio. A last correlated systematic uncertainty of $7.8\%$ takes into account the difference between the overall \jpsi/\Dz ratio, which is the average of the results integrated over $y^\star$ or \pt and multiplicity, and the individual results.

\begin{figure}[t]
\begin{center}
\includegraphics[width=0.49\textwidth]{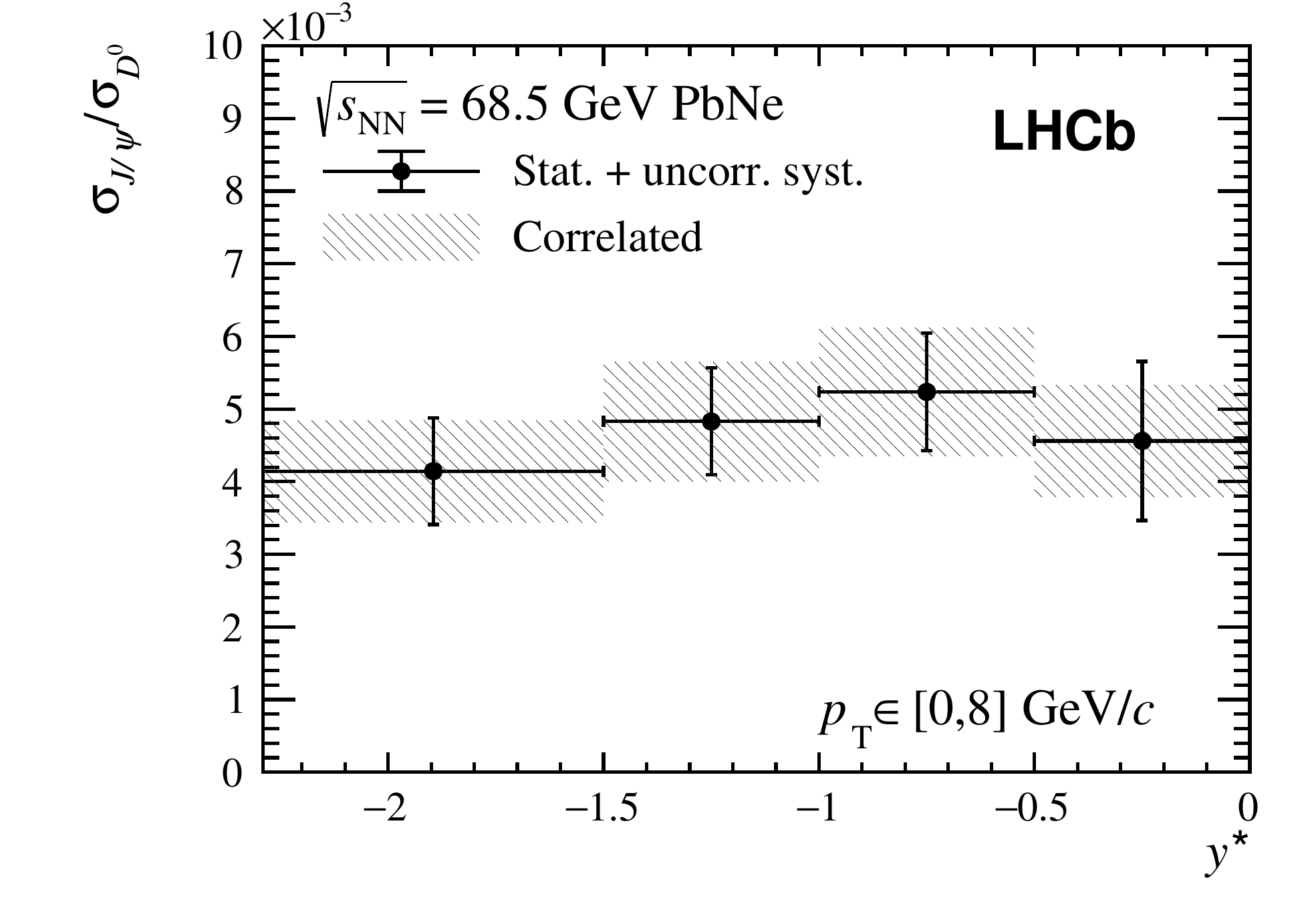}
\includegraphics[width=0.49\textwidth]{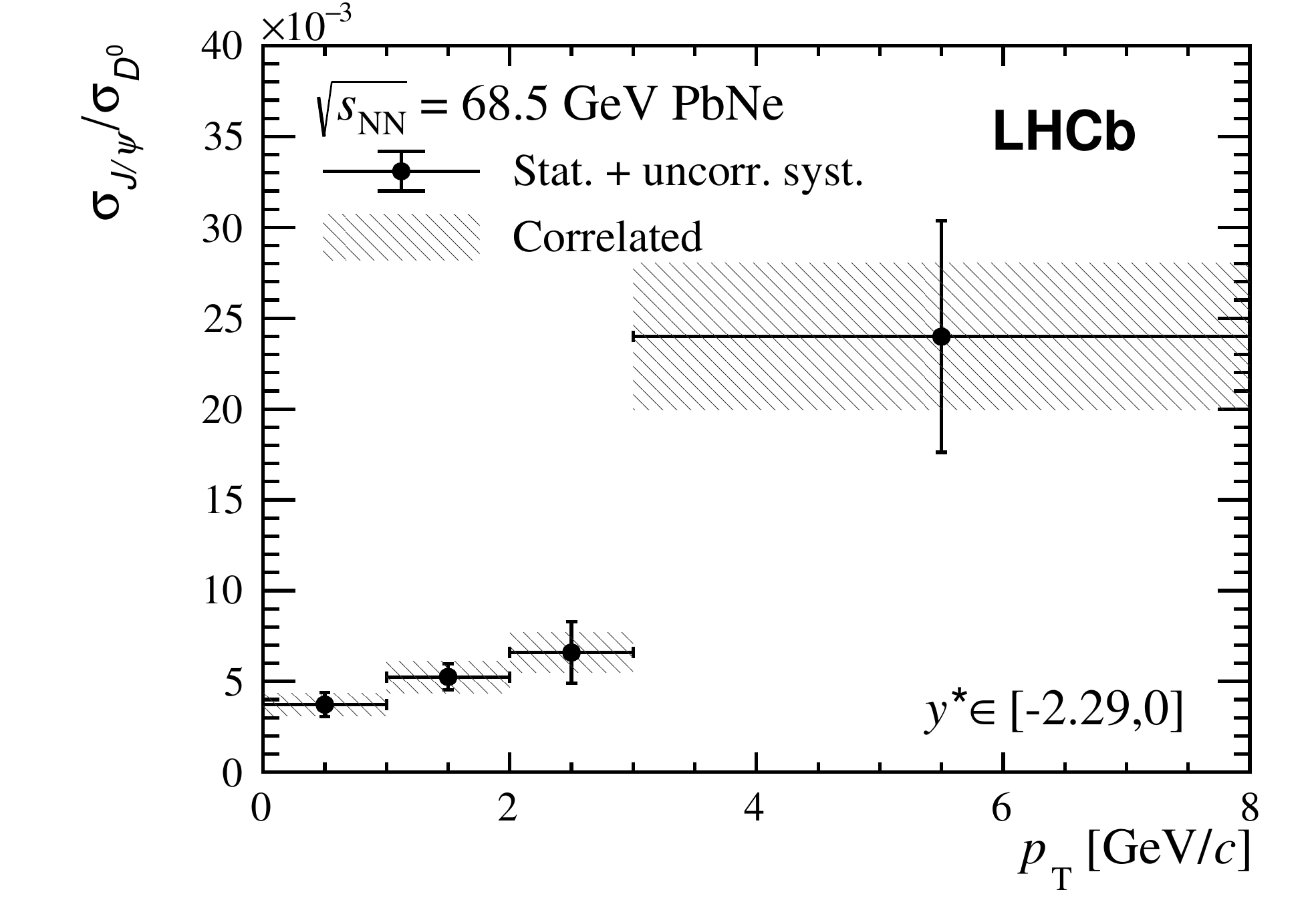}
\caption{Ratios of \jpsi to \Dz cross-sections as a function of (left) $y^\star$  and (right) transverse momentum. The error bars represent uncertainties that are uncorrelated bin-to-bin while the boxes represent the correlated uncertainties.}
\label{fig::psiD0}
\end{center}
\end{figure}

\begin{figure}
\begin{center}
\includegraphics[width=0.49\textwidth]{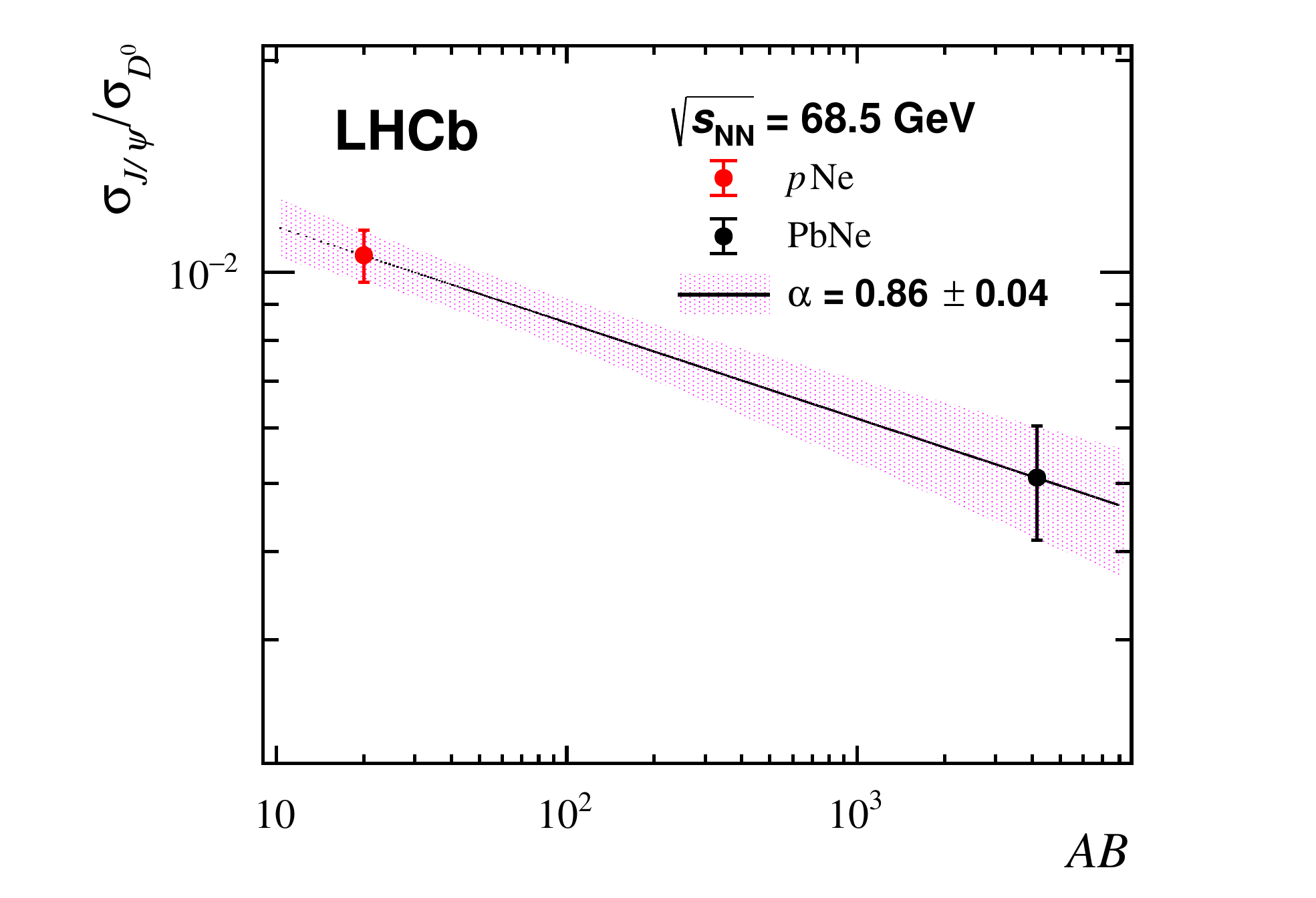} 
\includegraphics[width=0.49\textwidth]{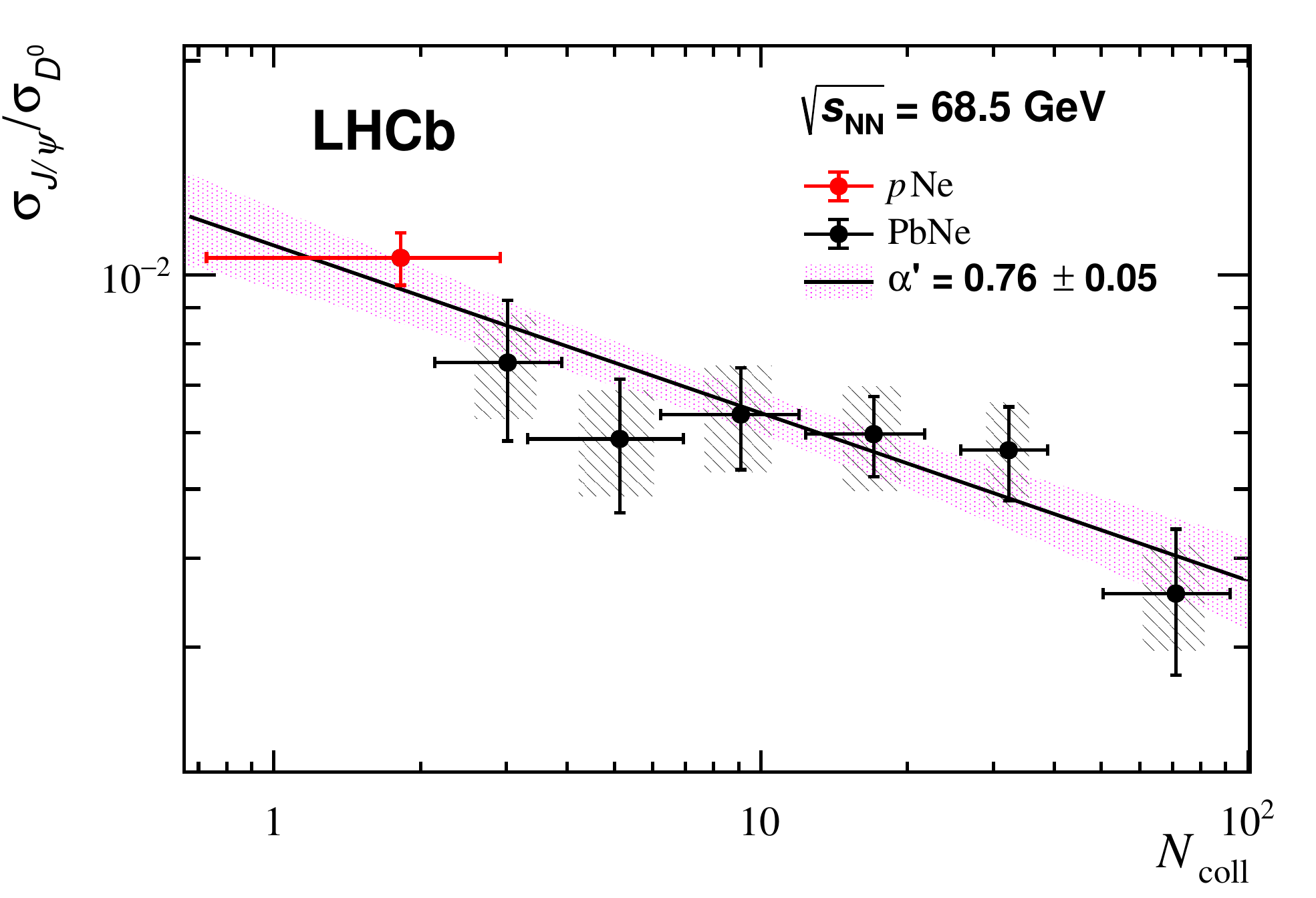} 
\caption{Left: \jpsi/\Dz cross-section ratio  as a function of $AB$, the product of the beam ($A$) and target ($B$) atomic mass numbers; the error bars represent the quadratic sum of statistical and systematic uncertainties. Right: \jpsi/\Dz cross-section ratio  as a function of \ncoll; the error bars represent the quadratic sum of statistical and uncorrelated systematic uncertainties while the boxes correspond to the correlated systematic uncertainty. The red and black points correspond to \pNe \cite{pNeJpsi} and PbNe collisions, respectively.} 
\label{fig::AB}
\end{center}
\end{figure}

%%%%% Results
The \jpsi/\Dz cross-section ratio, taking into account the branching fractions, is
\begin{equation*}
    \begin{aligned}
    \frac{\sigma_\jpsi}{\sigma_\Dz} &= \frac{Y_{\jpsi\to\mup\mun}}{\BF_{\jpsi\to\mup\mun}\times\varepsilon_\jpsi}\times\frac{\BF_{\Dz\to\Kmp\pipm}\times\varepsilon_\Dz}{Y_{\Dz\to\Kmp\pipm}}=0.51\pm0.04\stat \pm 0.10\syst \%,\\
    \end{aligned}
\end{equation*}
where $Y_{\jpsi\to\mup\mun}$ ($Y_{\Dz\to\Kmp\pipm}$), $\BF_{\jpsi\to\mup\mun}$ ($\BF_{\Dz\to\Kmp\pipm}$) and $\varepsilon_\jpsi$ ($\varepsilon_\Dz$) are \jpsi (\Dz) yield, branching fraction and total efficiency, respectively. The \jpsi/\Dz cross-section ratio as a function of $y^\star$ and \pt is shown in Fig.~\ref{fig::psiD0}. The ratio depends strongly on \pt but the data are compatible with no dependence on centre-of-mass rapidity. 
The \jpsi/\Dz cross sections ratios in PbNe collision data are compared with measurements in  \pNe collisions with \lhcb in similar experimental conditions \cite{pNeJpsi, pNeD0}. The left panel of Fig.~\ref{fig::AB} shows the \jpsi/\Dz cross-section ratio for both PbNe and \pNe data as a function of $AB$, the product of the beam ($A$) and target ($B$) atomic mass numbers. Assuming that nuclear effects, that can modify the initial \ccbar pair production, cancel in the $\jpsi/\Dz$ ratio, and that the \jpsi and \Dz production cross-sections are of the form $\sigma^{AB}_\jpsi=\sigma^{\pp}_\jpsi\times AB^\alpha$ and $\sigma^{AB}_\Dz=\sigma^{pp}_\Dz\times AB$, respectively, the cross-section ratio dependence on $AB$ is expected to follow
\begin{equation}
    \frac{\sigma_\jpsi^{AB}}{\sigma_\Dz^{AB}}=\frac{\sigma_\jpsi^{pp}}{\sigma_\Dz^{pp}}\times (AB)^{\alpha-1}=C\times (AB)^{\alpha-1}.
\end{equation}
A fit to the data gives $\alpha=0.86\pm0.04$ (and $C=(1.59\pm0.26)\times 10^{-2}$), which indicates that \jpsi production is affected by additional nuclear effects with respect to \Dz production. 

Figure \ref{fig::AB} (right) also shows the \jpsi/\Dz cross-section ratio as a function of the number of binary nucleon-nucleon collisions, \ncoll. The PbNe data sample is divided into intervals of \ncoll corresponding to different centrality intervals related to the overlap region between the two colliding nuclei. The larger the \ncoll the larger is the overlap region. Since small overlap regions correspond to small \ncoll values (similar to \ncoll values in proton-nucleus collisions), any suppression related to the formation of a deconfined medium should occur at large \ncoll values. 

The number of binary collisions is not directly observable but it can be mapped to the actual data using a Glauber model. In \lhcb, this mapping, reported in Table \ref{tab:cent_mapping}, is performed based on the hit multiplicity, as detailed in Ref. \cite{CentLHCb}. For \pNe collisions, the result is integrated over the impact parameter of the collision.
Assuming that the \jpsi and \Dz production cross-sections scale as $\mncoll^{\alpha'}$ and $\mncoll$ respectively, the cross-section ratio $\sigma_\jpsi/\sigma_{D^0}$ scales as $\mncoll^{\alpha'-1}$. A fit to the data (where the PbNe correlated systematic uncertainty is not taken into account) gives $\alpha'=0.76\pm0.05$, compatible with $\alpha$ when relating \ncoll to $AB$ with the Glauber model. This confirms that \jpsi meson production is affected by additional nuclear effects with respect to \Dz production. 
Moreover, within uncertainties, no difference in the \jpsi suppression trend is observed when comparing the PbNe largest \ncoll bin with the \pNe and PbNe smaller \ncoll bins.

\begin{table}[]
\centering
\caption{The mapping from intervals of the hit multiplicity to intervals of \ncoll; \mncoll and RMS(\ncoll) correspond to the \ncoll distribution mean and root mean square values respectively.}
\label{tab:cent_mapping}
\begin{tabular}{lccc}
\hline
                              & Hit multiplicity    & \mncoll         & RMS(\ncoll)     \\ 
\hline
{\textbf{\pNe}}               &             &  \phantom{0}1.81           &      \phantom{0}1.10         \\
\hline
{\textbf{PbNe}}&\phantom{00}0 -- 200\phantom{0} & \phantom{0}3.02 & \phantom{0}0.88        \\
&200 -- 300\phantom{0}           & \phantom{0}5.13 & \phantom{0}1.81       \\
&300 -- 446\phantom{0}           & \phantom{0}9.09 & \phantom{0}2.87      \\
&446 -- 715\phantom{0}           & 17.04           & \phantom{0}4.67       \\
&715 -- 960\phantom{0}           & 32.26           & \phantom{0}6.51        \\
&960 -- 1700                     & 71.12           & 20.70                  \\ 
\hline
\end{tabular}
\end{table}

In summary, we report on the \jpsi to \Dz production ratio in $\sqsnn=68.5\gev$ PbNe collisions with the \lhcb experiment. The \jpsi/\Dz cross-section ratio is found to be $\sigma_\jpsi/\sigma_\Dz = 0.51\pm0.04\stat\pm 0.10 \syst \%$ for $y^\star\in[-2.29, 0]$ and $\pt\in[0, 8]\gevc$. When compared with results obtained in \pNe collisions, the dependence of this ratio with $(AB)^{\alpha-1}$ gives $\alpha=0.86\pm0.04$. 
Finally, the study of this ratio, divided into intervals of \ncoll, shows no difference in the \jpsi suppression trend, when comparing \pNe and PbNe peripheral collisions with PbNe central collisions.

\section*{Acknowledgements}
%
% These Acknowledgements valid from 3-May-2019
%
\noindent We express our gratitude to our colleagues in the CERN
accelerator departments for the excellent performance of the LHC. We
thank the technical and administrative staff at the LHCb
institutes.
We acknowledge support from CERN and from the national agencies:
CAPES, CNPq, FAPERJ and FINEP (Brazil); 
MOST and NSFC (China); 
CNRS/IN2P3 (France); 
BMBF, DFG and MPG (Germany); 
INFN (Italy); 
NWO (Netherlands); 
MNiSW and NCN (Poland); 
MEN/IFA (Romania); 
%MSHE (Russia); 
MICINN (Spain); 
SNSF and SER (Switzerland); 
NASU (Ukraine); 
STFC (United Kingdom); 
DOE NP and NSF (USA).
We acknowledge the computing resources that are provided by CERN, IN2P3
(France), KIT and DESY (Germany), INFN (Italy), SURF (Netherlands),
PIC (Spain), GridPP (United Kingdom), 
%RRCKI and Yandex LLC (Russia), 
CSCS (Switzerland), IFIN-HH (Romania), CBPF (Brazil),
Polish WLCG  (Poland) and NERSC (USA).
We are indebted to the communities behind the multiple open-source
software packages on which we depend.
Individual groups or members have received support from
ARC and ARDC (Australia);
Minciencias (Colombia);
AvH Foundation (Germany);
EPLANET, Marie Sk\l{}odowska-Curie Actions and ERC (European Union);
A*MIDEX, ANR, IPhU and GLUODYNAMICS/Labex P2IO, and R\'{e}gion Auvergne-Rh\^{o}ne-Alpes (France);
Key Research Program of Frontier Sciences of CAS, CAS PIFI, CAS CCEPP, 
Fundamental Research Funds for the Central Universities, 
and Sci. \& Tech. Program of Guangzhou (China);
%Key Research Program of Frontier Sciences of CAS, CAS PIFI,
%Thousand Talents Program, and Sci. \& Tech. Program of Guangzhou (China);
%RFBR, RSF and Yandex LLC (Russia);
GVA, XuntaGal, GENCAT and Prog.~Atracci\'on Talento, CM (Spain);
SRC (Sweden);
the Leverhulme Trust, the Royal Society
 and UKRI (United Kingdom).

\clearpage

\addcontentsline{toc}{section}{References}
\setboolean{inbibliography}{true}

\ifx\mcitethebibliography\mciteundefinedmacro
\PackageError{LHCb.bst}{mciteplus.sty has not been loaded}
{This bibstyle requires the use of the mciteplus package.}\fi
\providecommand{\href}[2]{#2}

\clearpage
% LHCb collaboration author list
% Data extracted on June 1st, 2023 at 11:40am for paper reference LHCb-PAPER-2022-011
\centerline
{\large\bf LHCb collaboration}
\begin
{flushleft}
\small
R.~Aaij$^{32}$\lhcborcid{0000-0003-0533-1952},
A.S.W.~Abdelmotteleb$^{50}$\lhcborcid{0000-0001-7905-0542},
C.~Abellan~Beteta$^{44}$,
F.~Abudin{\'e}n$^{50}$\lhcborcid{0000-0002-6737-3528},
T.~Ackernley$^{54}$\lhcborcid{0000-0002-5951-3498},
B.~Adeva$^{40}$\lhcborcid{0000-0001-9756-3712},
M.~Adinolfi$^{48}$\lhcborcid{0000-0002-1326-1264},
H.~Afsharnia$^{9}$,
C.~Agapopoulou$^{13}$\lhcborcid{0000-0002-2368-0147},
C.A.~Aidala$^{76}$\lhcborcid{0000-0001-9540-4988}
% ...

{\footnotesize \it

% ...
$^{9}$Universit{\'e} Clermont Auvergne, CNRS/IN2P3, LPC, Clermont-Ferrand, France\\
% ...
$^{13}$LPNHE, Sorbonne Universit{\'e}, Paris Diderot Sorbonne Paris Cit{\'e}, CNRS/IN2P3, Paris, France\\
% ...
$^{32}$Nikhef National Institute for Subatomic Physics, Amsterdam, Netherlands\\
% ...
$^{40}$Instituto Galego de F{\'\i}sica de Altas Enerx{\'\i}as (IGFAE), Universidade de Santiago de Compostela, Santiago de Compostela, Spain\\
% ...
$^{44}$Physik-Institut, Universit{\"a}t Z{\"u}rich, Z{\"u}rich, Switzerland\\
% ...
$^{48}$H.H. Wills Physics Laboratory, University of Bristol, Bristol, United Kingdom\\
% ...
$^{50}$Department of Physics, University of Warwick, Coventry, United Kingdom\\
% ...
$^{54}$Oliver Lodge Laboratory, University of Liverpool, Liverpool, United Kingdom\\
% ...
$^{62}$Syracuse University, Syracuse, NY, United States\\
% ...
$^{76}$University of Michigan, Ann Arbor, MI, United States, associated to $^{62}$\\
\bigskip
% ...
}
\end{flushleft}

\end{document}